\pdfoutput=1
\documentclass[10pt,twocolumn,letterpaper]{article}

\usepackage{iccv}
\usepackage{times}
\usepackage{epsfig}
\usepackage{graphicx}
\usepackage{amsmath}
\usepackage{amssymb}
\usepackage[ruled,vlined]{algorithm2e}
\usepackage{algorithmic}
\usepackage{multirow}
\usepackage{color,soul}
\newcommand{\ignore}[1]{}
\usepackage{tabularx}
\newcommand{\uptox}{48.5}

\renewcommand\texthl[1]{#1}

\usepackage[pagebackref=true,breaklinks=true,letterpaper=true,colorlinks,bookmarks=false]{hyperref}

\iccvfinalcopy 

\def\iccvPaperID{2756} 

\begin{document}

\title{Enabling Inference Privacy with Adaptive Noise Injection}

\author{Sanjay Kariyappa\\
Georgia Institute of Technology\\
Atlanta GA, USA\\
{\tt\small sanjaykariyappa@gatech.edu}
\and
Ousmane Dia\\
Facebook\\
Menlo Park CA, USA\\
{\tt\small ousamdia@fb.com}
\and
Moinuddin K Qureshi\\
Georgia Institute of Technology\\
Atlanta GA, USA\\
{\tt\small moin@gatech.edu}
}

\maketitle

\ignore{

Flow

1.   

}

\begin{abstract}

   User-facing software services are becoming increasingly reliant on remote servers to host Deep Neural Network (DNN) models, which perform inference tasks for the clients. Such services require the client to send input data to the service provider, who processes it using a DNN and returns the output predictions to the client. Due to the rich nature of the inputs such as images and speech, the input often contains more information than what is necessary to perform the primary inference task. Consequently, in addition to the primary inference task, a malicious service provider could infer secondary (sensitive) attributes from the input, compromising the client’s privacy.  The goal of our work is to improve inference privacy by injecting noise to the input to hide the irrelevant features that are not conducive to the primary classification task. To this end, we propose Adaptive Noise Injection (ANI), which uses a light-weight DNN on the client-side to inject noise to each input, before transmitting it to the service provider to perform inference. Our key insight is that by customizing the noise to each input, we can achieve state-of-the-art trade-off between utility and privacy (up to $\uptox \%$ degradation in sensitive-task accuracy with $<1\%$ degradation in primary accuracy), significantly outperforming existing noise injection schemes. Our method does not require prior knowledge of the sensitive attributes and incurs minimal computational overheads.
\end{abstract}

\section{Introduction}

The unprecedented success of deep learning in recent years has led to its growing adoption in a number of products and applications. Additionally, Machine Learning (ML) inference is being offered as a service by several companies to perform various inference tasks for the client. Typically, the state-of-the-art models used in these services are computationally demanding, involving on the order of billions of parameters and trillions of FLOPs per inference (For instance, the recently proposed GPT-3 language model~\cite{gpt3} has around 175 billion parameters). The high memory and compute requirements prevent the deployment of these models on resource-constrained edge devices such as smartphones and smart home devices. 
Instead, DNN models are often hosted as remote services, wherein the client sends the input data to a remote server hosting the DNN model, which performs inference on the input and returns the predictions (or some action based on the prediction) to the client. Voice assistants (Siri, Alexa, Google Assistant) and computer vision services (Google Vision AI, Clarifi) are examples of such cloud-based ML services. Unfortunately, since the inputs sent by the client typically contain more information than what is necessary for the primary classification task, a malicious service provider can infer additional sensitive attributes from the input, compromising the privacy of the client.

\begin{figure}[b]
	\centering
    \centerline{\epsfig{file=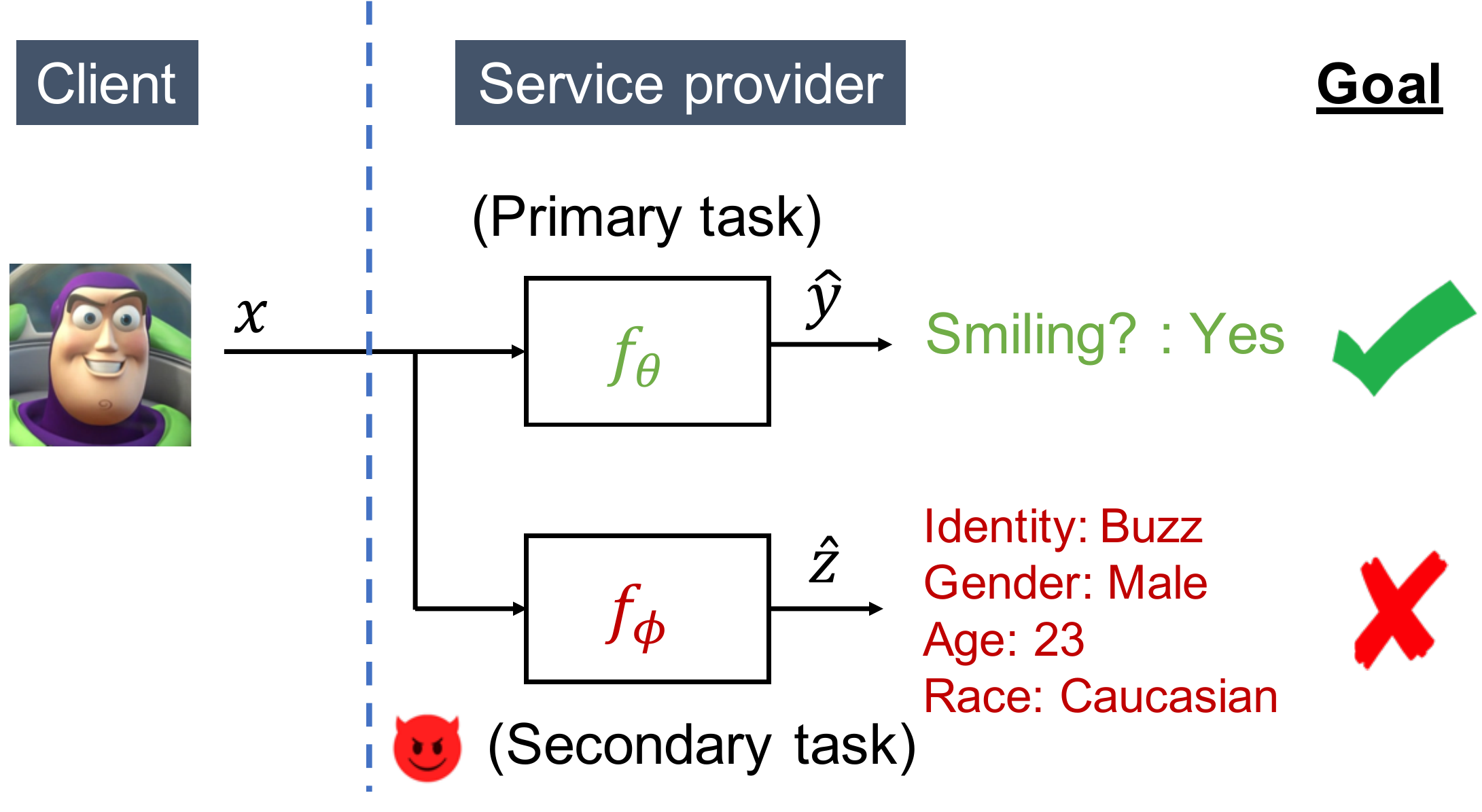, width=0.9\columnwidth}}
	\caption{In addition to the primary classification task: $\hat{y}=f_\theta(x)$, a  malicious service provider can reuse the same input to perform secondary classification tasks: $\hat{z}=f_\phi(x)$, which compromises the client's privacy by revealing sensitive attributes of the input.} 
    \label{fig:motivation}
\end{figure}

To explain, consider a service provider hosting a smile detection service on a remote server as shown in Fig.~\ref{fig:motivation}. A client availing this service is required to send the picture of a person and the service provider returns a prediction indicating if the person in the picture is smiling. Note that the client has no control over the input image once it has been sent to the service provider. Consequently, in addition to the primary task (smile detection), a malicious service provider could choose to perform secondary classification tasks on the same input that reveal possibly sensitive information such as the gender, race, age and, identity of the person in the input image, compromising the client's privacy.


The goal of our work is to improve inference privacy in remote ML inference settings. Fundamentally, the reason why the service provider is able to infer sensitive attributes is that the unmodified input $x$ that is sent to the service provider contains more information than what is needed to perform the primary classification task.  This paper proposes \emph{Adaptive Noise Injection} (ANI) to improve input privacy by transforming the input $x$ to $x'$ on the client-side by adaptively injecting noise to the input. ANI injects noise in a way that allows the service provider to infer the primary attributes but deters the inference of secondary attributes from the transformed input $x'$. Varying the amount of noise injected allows the client to trade-off utility (high accuracy on primary task) for privacy (low accuracy on secondary task). 

A similar insight has been used by recent work \emph{Cloak}~\cite{Cloak} to improve inference privacy, \texthl{which uses a fixed noise distribution to perturb the input and hide the non-conducive features. However, the set of conducive and non-conducive features are not fixed and can change significantly across various inputs. For instance, there can be a large spatial variation in the pixels corresponding to the mouth (conducive features) in the smile-detection task. Consequently, Cloak's ability to obfuscate non-conducive features (to provide better privacy) is limited by the need to capture this variance in the conducive features of the input, which cannot be done efficiently using a fixed noise distribution. Our proposal ANI overcomes this limitation by developing a framework that allows for a variable amount of noise injection, which enables the noise to be tailored to each input. This flexibility allows ANI to obfuscate a greater fraction of the non-conducive features, providing a significantly better trade-off between utility and privacy compared to Cloak.} We make the following key contributions in this paper:

\begin{itemize}
\item We develop a noise-injection framework called \emph{Adaptive Noise Injection} (ANI) to improve input privacy in remote ML inference settings. Our proposal leverages a custom neural network to inject noise that is tailored to each input, before sending it to the service provider for remote inference.

\item We propose a method to train the noise-injection network that balances the utility (high accuracy on the primary task) and privacy (low accuracy on secondary tasks) objectives, without prior knowledge of the sensitive secondary attributes  that need to be protected. We provide theoretical analysis that connects our training objective with the mutual information terms corresponding to the input and the class labels.
\item We evaluate ANI over multiple image classification tasks defined on MNIST, FashionMNIST, CIFAR-100, and CelebA datasets to demonstrate that, owing to the input-specific nature of our noise injection scheme, ANI provides a state-of-the-art trade-off between utility and privacy (up to $\uptox \%$ degradation in sensitive-task accuracy with $<1\%$ degradation in primary accuracy), significantly outperforming prior work.
\end{itemize}
\ignore{
Hardware and algorithmic approaches have been developed to address the problem of input privacy in remote ml inference. We provide a brief overview of related works and discuss the key drawbacks that limit their adoption.

}
\section{Related Work}
Several recent works have been developed both in the hardware and machine learning communities to address the problem of input privacy in remote ML inference. Prior works can be broadly divided into hardware-based and algorithmic solutions. In this section, we provide a brief overview of recent related works and discuss the key drawbacks that limit their adoption.

\subsection{Hardware-based Solutions}
\emph{Trusted Execution Environments} (TEEs) such as Intel SGX~\cite{sgx} and ARM TrustZone~\cite{trustzone} provide hardware support to execute code remotely with confidentiality and integrity. TEEs provide attestation that guarantees the correct execution of the code (integrity). In addition, it uses encryption and software support to isolate the data and the code from other processes, operating system, and the hardware owner, protecting the confidentiality of the data. These features allow the remote execution of DNNs on TEEs while ensuring the privacy of the input data from the cloud service provider. Unfortunately, commercially available TEEs systems are limited to CPUs and do not support trusted execution on GPUs, limiting throughput ($>50\times$ lower throughput compared to GPU~\cite{slalom}). \texthl{Additionally, TEEs have limited memory capacity and poor memory performance, which are detrimental for large-scale DNN workloads.}
These factors result in TEEs being orders of magnitude slower than untrusted alternatives like GPUs. User-facing ML services usually have stringent latency requirements for processing the inference request to ensure quality-of-service, making TEEs ill-suited for such latency-sensitive applications. A recent line of work has investigated using a combination of trusted and untrusted hardware (TEE+GPUs) to improve execution times by offloading most of the compute to GPUs~\cite{slalom, darknight}. While these proposals do provide speedup compared to TEE-only execution, they are still considerably slower compared to GPUs in terms of inference latency and throughput. Furthermore, hardware-based solutions require special-purpose CPUs that support TEEs, which adds to the cost.

\subsection{Algorithmic Solutions}
Several algorithmic solutions have also been proposed to enable inference privacy using non-secure hardware. They can be broadly classified into Homomorphic Encryption (HE) and input-preprocessing-based solutions.

\textbf{Homomorphic Encryption}:
HE allows DNN computations to be performed on encrypted data. Since the input and the resulting output are both in the encrypted form, this protects the privacy of the input data from the service provider. 
Unfortunately, HE incurs significant computational overheads, resulting in around 5-6 orders of magnitude increase in execution times compared to unencrypted inference~\cite{cryptonets, cryptodl}. Additionally, HE does not natively support the evaluation of  non-linearities like ReLU and MaxPool. Existing proposals have tried to circumvent this problem either using polynomial approximations of non-linear layers~\cite{lola,cryptodl, cryptonets} or by relying on the client to evaluate the non-linear layers~\cite{gazelle, cheetah} using garbled circuits~\cite{gc}. Unfortunately, both of these solutions have their own set of drawbacks. Polynomial approximations cause degradations in accuracy and impose large computational overheads. Relying on the client to compute the non-linear operations, on the other hand, requires multiple rounds of communication between the client and server, adding to the latency of inference. 
A recent line of work~\cite{gazelle, cheetah} has shown significant speedups in inference time through a combination of kernel-level optimizations, changes to HE parameters, and custom hardware. However, even with these optimizations, HE algorithms fall short of real-time inference requirements.

\textbf{Input-preprocessing:}
Instead of sending the unmodified input to the server, the client can transform the input $x$ to $x'$ and send this transformed representation to the server to perform inference. The goal of this preprocessing is to remove the irrelevant information in the input that is not conducive to the primary classification task. Our work falls under this category. We discuss two recent input-preprocessing based solutions and describe the limitations of these prior works.~\footnote{In addition to the related works in this section, we also discuss other input-preprocessing based proposals in Appendix~\ref{app:other_related_works}.}

\emph{1. Deep Private Feature Extraction} (DPFE)~\cite{dpfe}: DPFE trains a transformation function $g$ such that, given a pair of points $(x_i, x_j)$, it maximizes the distance between the transformed representations $\left\|x'_{i}-x'_{j}\right\|_{2}$ if the inputs belong to the same secondary (sensitive) class and minimizes this distance if the inputs belong to different secondary classes. This yields the following optimization objective:
\begin{align}~\label{eqn:dpfe}
   \min_{g} \sum_{(i, j): z_{i} \neq z_{j}}\left\|x'_{i}-x'_{j}\right\|_{2}^{2}-\sum_{(i, j): z_{i}=z_{j}}\left\|x'_{i}-x'_{j}\right\|_{2}^{2}.
\end{align}
By performing this optimization, DPFE makes it harder for an adversary to infer the secondary attributes  using the transformed representations $x'$. Unfortunately, the optimization in Eqn.~\ref{eqn:dpfe} requires prior knowledge of the sensitive attributes $\{z\}$ of the training data, which can be problematic for several reasons. First, this scheme requires the service provider and the end-user to agree on a list of sensitive attributes that may not be exhaustive, leaving open the possibility of discovering additional sensitive attributes in the future. Second, the service provider needs to collect the sensitive labels corresponding to the training data, which can be expensive. To avoid these issues, our work assumes that the sensitive attributes corresponding to the input are not known to the client a priori.

\emph{2. Cloak}: Similar to our proposal, a recent work \emph{Cloak}~\cite{Cloak} proposes to preprocess the input by injecting noise with the goal of hiding non-conducive features as shown in the equations below:
 \begin{align}~\label{eqn:Cloak}
   x' = (x +v)\cdot b +\mu,\\
   v \sim \mathcal{N}(0, \sigma^2).
\end{align}
The parameters $\sigma$ and $\mu$ are learnable tensors that are learned during the training process. The training objective of Cloak aims to maximize $\sigma$, while retaining the accuracy of the primary task. \texthl{A key limitation of Cloak is that it uses the same noise parameters ($\mu, \sigma$) to inject noise for all the inputs, which limits its efficacy. In contrast, we propose a framework that allows the noise distribution to be customized for each input. This enables ANI to modify the noise to account for the variations of conducive/non-conducive features in the input and obfuscate a greater fraction of the non-conducive features compared to Cloak, providing better privacy.} We treat Cloak as our baseline and through extensive empirical analysis show that owing to the input-specific nature of noise injection, ANI provides a significantly better utility-privacytrade-off compared to Cloak.\footnote{Note that Cloak was originally proposed in the setting where the service provider's model is fixed. Since our work involves retraining the service provider's model, we also retrain the service provider's model for Cloak to make the comparison fair.}

\emph{}

\section{Preliminaries}
The goal of our work is to enable input privacy for the client in a remote ML inference setting by preprocessing the input through adaptive noise injection. Before discussing our solution, we provide context by formalizing the problem setting and outlining the objectives of input-preprocessing-based solutions.

\subsection{Problem Setup}~\label{sec:problem_setup}
Our problem setup involves a client who interacts with a service provider to perform inference on the client's input data $\{x_i\}_i$ . Let $\mathcal{D}_{test}=\{x_i,y_i,z_i\}_i$ denote the client's input data and the corresponding ground truth labels. Here, $y_i$ and $z_i$ denote the ground truth primary and secondary (sensitive) labels of the input data respectively. Similarly, let $\mathcal{D}_{train}=\{x_j,y_j,z_j\}_j$ denote the service provider's training data. The client and the service provider have the following objectives and constraints.

\textbf{Common Utility Objective:}  The client and service provider share the utility objective of performing inference on the primary task. The service provider uses a model $f_\theta$, trained on $\mathcal{D}_{train}$ to provide predictions of the primary labels: $\hat{y} = f_\theta(x)$ for samples from $\mathcal{D}_{test}$ to the client.

\textbf{Opposing Privacy Objectives:}  The client and service provider have opposing privacy objectives with regards to the secondary task. The service provider can train a secondary model $f_\phi$ on $\{x_j,z_j\}_j$ to obtain predictions of the secondary attribute $\hat{z} = f_\phi(x)$ on the client's input data, compromising the client's input privacy. On the other hand, the client is interested in protecting input privacy by preventing the service provider from inferring the secondary attributes associated with the client's inputs. 

\textbf{Constraints on Secondary Labels:} In this work, we assume that the secondary labels $z$ corresponding to the data is not known to the client a priori. We argue that this problem setting is more relevant to real-world applications, where it is hard to exhaustively list the sensitive attributes and collect the corresponding labels for large datasets.



\subsection{Enabling Privacy through Input-Preprocessing}\label{sec:utility_privacy}
 Our goal is to enable inference privacy by transforming the input $x$ to $x'$ on the client-side and sending this transformed representation of the input $x'$ to the service provider for inference. Let $X, X', Y$ and $Z$ denote the random variables that represent the input ($x$), transformed input ($x'$), primary ($y$) and secondary labels ($z$) respectively. We want the input transformation to meet two key objectives:

\begin{enumerate}
\item \textit{Utility objective:} The service provider should be able to predict the primary label for the client's inputs i.e. the model used for the primary task $f_\theta$ should have high accuracy. Thus, the input transformation should preserve the mutual information between $X'$ and $Y$.
\item \textit{Privacy objective:} The service provider should not be able to infer the secondary labels from $x'$. In other words, the mutual information between $X'$ and $Z$ must be minimized.
\end{enumerate}

By combining the privacy and utility objectives, finding the optimal $X'$ can be viewed as an optimization problem in terms of the mutual information ($\mathcal{I}$) between $X', Y$ and $Z$ as shown below:

\begin{align}~\label{eq:MI_optimization}
    \min \lambda\mathcal{I}(X'; Z) -  \mathcal{I}(X'; Y).
\end{align}

Here, $\lambda$ denotes the relative importance between the utility and privacy objectives. In the following section, we describe the design of our proposal \emph{Adaptive Noise Injection}, which uses a DNN $f_\psi$ to produce $x'$ and describe a method to jointly train $f_\psi$ and $f_\theta$ by balancing the utility and privacy objectives. We also derive a theoretical connection between the optimization objective in Eqn.~\ref{eq:MI_optimization} and the loss function used to train our models.

\section{Adaptive Noise Injection}
We propose \emph{Adaptive Noise Injection} (ANI) to enable inference privacy. ANI generates a noisy representation of the input $x'$ on the client-side by hiding the features that are not conducive to the primary inference task. $x'$ can then be transmitted to the service provider to perform inference for the primary task. By hiding irrelevant features, our proposal deters the service provider's ability to infer the secondary attributes corresponding to the input, improving privacy. We start by describing the design of our noise injection mechanism that leverages a neural network ($f_\psi$) to tailor the noise for each input. The noise injection mechanism exposes a trade-off between utility and privacy as described in Eqn.~\ref{eq:MI_optimization}. With this equation as the starting point, we derive a loss function, which can be used to jointly train the noise injection network (deployed by the client) and the primary classification network (deployed by the service provider) using gradient descent.   

\subsection{Design}
\begin{figure}[htb]
 \vspace{-0.05 in}
	\centering
    \centerline{\epsfig{file=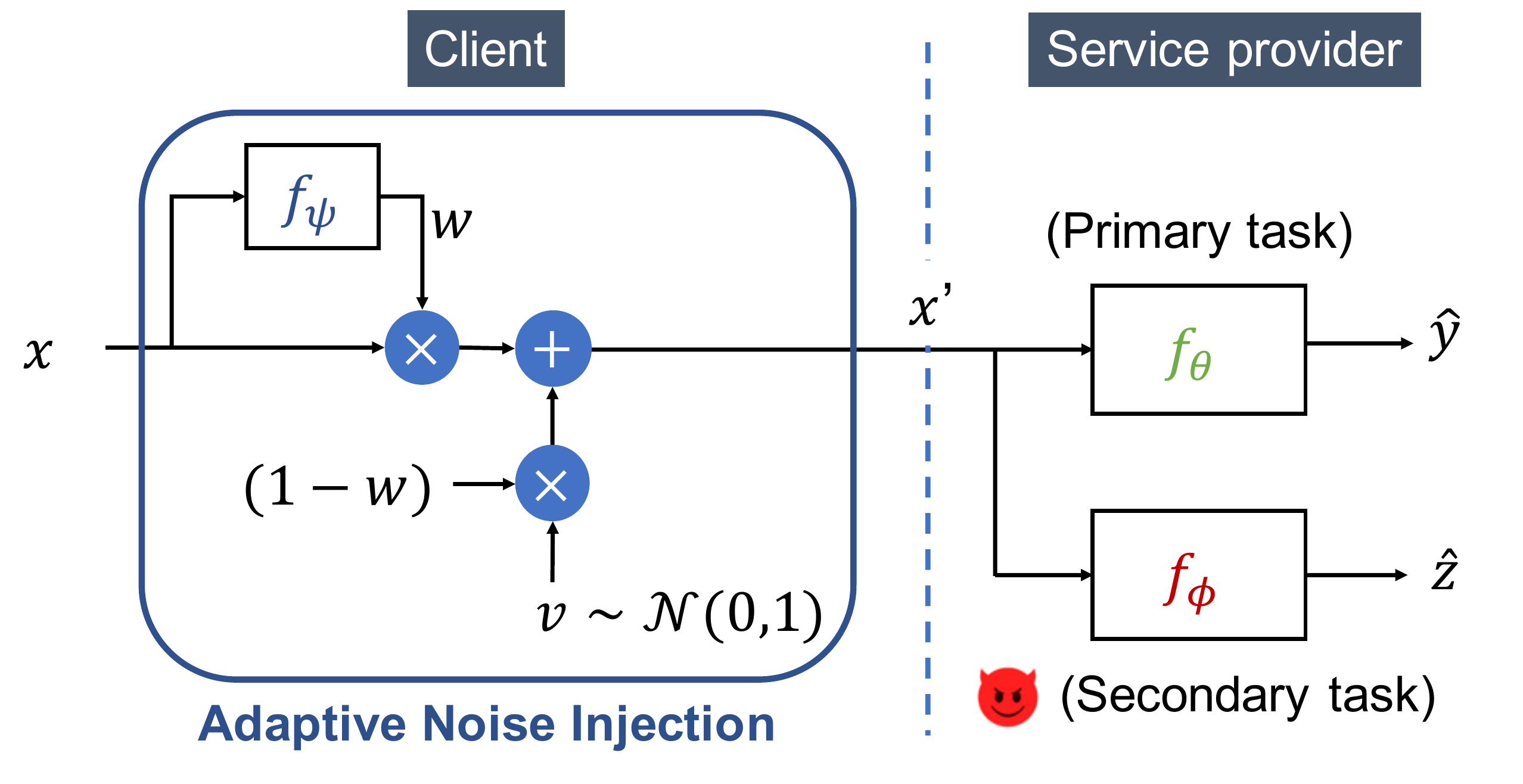, width=\columnwidth}}
	\caption{Design of \emph{Adaptive Noise Injection} (ANI). ANI injects noise to the input $x$ to produce a transformed representation $x'$, which is sent to the service provider for inference. ANI uses a neural network $f_\psi$ to tailor the noise for each input.}
    \label{fig:ani_design}
    
\end{figure}

The design of our proposed noise injection scheme ANI is shown in Fig.~\ref{fig:ani_design}. Given an input $x\in \mathbb{R}^N$, ANI produces a noisy representation $x'\in \mathbb{R}^N$ by taking a weighted average of $x$ and a noise vector $v\in \mathbb{R}^N$, sampled from a unit normal distribution, as shown in the equation below:
\begin{align}~\label{eq:x'}
    x' = w\circ x+(1-w)\circ v \\
    \text{where, }v \sim\mathcal{N}(0,1); w\in [0,1]^N.
\end{align}

The weight vector $w$ determines the amount of noise injected to the inputs. ANI uses a neural network $f_\psi$ to generate the weights as a function of the input: $w=f_\psi(x)$. Owing to the input-specific nature of the weights, ANI provides the flexibility to inject a variable amount of noise customized to each input. Next, we derive a loss function that can be used to jointly train $f_\psi$ and $f_\theta$ by balancing utility and privacy.

\subsection{Derivation of the Loss Function}

We are interested in deriving a loss function that can be used to train $f_\psi$ and $f_\theta$. Let $\psi$ and $\theta$ denote the parameters of $f_\psi$ and $f_\theta$ respectively. We start by considering the following optimization objective over the network parameters $\psi$, which is used by $f_\psi$ to produce $x'$ by balancing utility and privacy (as discussed in Section~\ref{sec:utility_privacy}): 

\begin{align}~\label{eq:MI_optimization2}
    \min_{\psi} \lambda\mathcal{I}(X'; Z) -  \mathcal{I}(X'; Y).
\end{align}

Since we assume that the sensitive attributes $Z$ are not known to the user a-priori, we re-express the above objective function by replacing $\mathcal{I}(X'; Z)$ with $\mathcal{I}(X'; X)$ using the following theorem.

\textit{Theorem 1.} Given random variables  $X, X'$ and $Z$ where $X' = f(X)$, $\mathcal{I}(X'; Z)$ has the following upper-bound:

\begin{align}
\mathcal{I}(X'; X) \geq \mathcal{I}(X'; Z).
\end{align}
\textit{Proof.} See Appendix~\ref{app:info_xy}.

Based on Theorem 1, we derive the following upper-bound for Eqn.~\ref{eq:MI_optimization2}:
\begin{equation}\label{eq:upper_bound}
    \lambda \mathcal{I}(X'; X) - \mathcal{I}(X'; Y) \geq \lambda \mathcal{I}(X'; Z) - \mathcal{I}(X'; Y).
\end{equation}

Minimizing $\mathcal{I}(X'; X)$ in the above equation can be viewed as maximizing the conditional entropy $H(X'|X)$. To explain, we start by noting that $\mathcal{I}(X'; X)$ can be expressed as follows:

\begin{equation}
    \begin{gathered} \label{eq:h0}
       \mathcal{I}(X'; X) \equiv H(X') - H(X'|X).
    \end{gathered}
\end{equation}

 As we increase $H(X'|X)$ by reducing $w$ in Eqn.~\ref{eq:x'}, $H(X'|X)\rightarrow H(X')$ and $\mathcal{I}(X'; X)\rightarrow 0$. Thus, by replacing $\mathcal{I}(X'; X)$ by $-H(X'| X)$ in Eqn.~\ref{eq:upper_bound}, we have the following optimization objective:
 
\begin{equation} \label{eq:h1}
       \min_{\psi} -\lambda H(X'| X) - \mathcal{I}(X';Y).
\end{equation}

\textbf{Estimating $\boldsymbol{\mathcal{I}(X';Y)}$:} Consider the function $f_\theta: \mathcal{X} \rightarrow \mathcal{Y}$. Let $\max_{\theta}\mathcal{I}(f_\theta(X'); Y)$ denote the maximum mutual information that can be obtained by optimizing over $\theta$. We have the following bounds for $\mathcal{I}(X'; Y)$:

\begin{align}
       \max_{\theta}\mathcal{I}(f_\theta(X'); Y) &\leq \mathcal{I}(X'; Y)\\
       -\max_{\theta}\mathcal{I}(f_\theta(X'); Y) &\geq -\mathcal{I}(X'; Y)\\
       \min_{\theta}(-\mathcal{I}(f_\theta(X'); Y)) &\geq -\mathcal{I}(X'; Y).~\label{eq:lower_bound}
\end{align}

Substituting the upper bound of Eqn.~\ref{eq:lower_bound} in Eqn.~\ref{eq:h1}, we obtain the following:

\begin{align} 
    \min_{\psi} \big(-\lambda H(X'| X) &+ \min_{\theta}-\mathcal{I}(f_\theta(X');Y)\big)\\
   \min_{\psi, \theta} -\lambda H(X'| X) &- \mathcal{I}(f_\theta(X'); Y).\label{eq:h2}
\end{align}

Recent work~\cite{boudiaf2020metric} has shown that maximizing the mutual information is equivalent to minimizing the cross entropy ($\mathcal{H}$) between random variables. Thus, we can replace $\min_{\psi, \theta}-\mathcal{I}(f_\theta(X'); Y)$ with $min_{\psi, \theta} \mathcal{H}(Y; f_\theta(X'))$ in Eqn.~\ref{eq:h2}, which gives:

\begin{equation}
    \begin{gathered} \label{eq:h3}
       \min_{\psi, \theta} -\lambda H(X'| X) + \mathcal{H}(Y; f_\theta(X')).
    \end{gathered}
\end{equation}

\textbf{Estimating $\boldsymbol{H(X'| X)}$:} From Eqn.~\ref{eq:x'}, we have $X' = w X + (1-w)V$, where $V$ is a random variable that follows a standard normal distribution. If $X$ is known, all the entropy in $X'$ comes from $V$, which gives us the following:

\begin{align} 
    H(X'|X) &= H((1-w)\cdot V)\\
    H(X'|X) &= 0.5\ \ln\det(2\pi e \Sigma).\label{eq:h4}
\end{align}

Here, $\Sigma$ is a diagonal covariance matrix with $1-w_i$ making up the diagonal elements. Eqn.~\ref{eq:h4} can be expressed in terms of the weights $w_i$ as shown below:

\begin{align}
    H(X'|X) &= 0.5\ \ln \big((2\pi e)^N \det(\Sigma)\big)\\
    H(X'|X) &= 0.5\ \ln \big((2\pi e)^N \Pi_i(1-w_i)\big)\\
    H(X'|X) &= 0.5\ \ln(2\pi e)^N + 0.5 \sum_{i=0}^{N-1}\ln(1-w_i).  \label{eq:h5}
\end{align}

Ignoring the constant terms in Eqn.~\ref{eq:h5} and substituting it in Eqn.~\ref{eq:h3}, we get the following optimization objective:
\begin{equation}
    \begin{gathered} \label{eq:h6}
       \min_{\psi, \theta} -\lambda \sum_{i=0}^{N-1}\ln(1-w_i) + \mathcal{H}(Y; f_\theta(X')).
    \end{gathered}
\end{equation}

\textbf{Final Loss Function:} Given a dataset $\mathcal{D}$, we get the following loss function, which can be used to perform gradient descent over model parameters $\theta$ and $\psi$:

\begin{align} \label{eq:loss_fn}
\mathcal{L} &= \mathbb{E}_{(x,y)\sim \mathcal{D}}\mathcal{H}(y; \hat{y})  - \lambda \sum_{i=0}^{N-1}\ln(1-w_i)\\
&\text{ where }, \hat{y} = f_\theta(x'); w = f_\psi(x).\nonumber
\end{align}

Intuitively, the first term in our loss function tries to maximize utility by minimizing the cross entropy loss between the predicted $\hat{y}$ and the true labels $y$. The second term encourages the weight values to be lower in Eqn.~\ref{eq:x'}, which increases the amount of noise injected, improving privacy. 


                
\begin{algorithm} ~\label{alg:training}
\DontPrintSemicolon 

\KwIn{$\mathcal{D}_{train} = \{x_i, y_i\}, \lambda \text{, learning rate: }\eta$}
\KwOut{Noise Injection model $f_\psi(\cdot;\psi)$ and Primary task model $f_\theta(\cdot;\theta)$}
Initialize $f_\psi(\cdot;\psi), f_\theta(\cdot;\theta)$\;
\Repeat{Convergence}{
\For{$\{x, y\}_\text{batch}$ in $\mathcal{D}_\text{train}$} {
    \For{$(x_i, y_i)$ in $\{x, y\}_\text{batch}$} {
        $v \sim \mathcal{N}(0,I_N)$\;
        $w = f_\psi(x_i)$\;
        $x'_i = x_i\circ w + (1-w)\circ v$\;
        $\hat{y}_i = f_\theta(x'_i)$\;
        $\mathcal{L}_i = \mathcal{H}(y_i; \hat{y}_i) -\lambda \sum_i\ln(1-w_i)$\;
    }
    $\mathcal{L} = \mathbb{E}(\mathcal{L}_i)$\;
    $\psi \gets \psi - \eta\cdot \nabla_{\psi}\mathcal{L}$\;
    $\theta \gets \theta - \eta\cdot \nabla_{\theta}\mathcal{L}$\;
}
}
\caption{ANI training algorithm}
\end{algorithm}

\subsection{Training Algorithm}
Algorithm~\ref{alg:training} shows how the loss function in Eqn.~\ref{eq:loss_fn} can be used to jointly train $f_\theta$ and $f_\psi$ using gradient descent with the training data $\mathcal{D}_{train}$. The hyperparameter $\lambda$ controls the relative importance between the utility and privacy objectives. A higher value of $\lambda$ provides a higher level of privacy at the cost of lower accuracy on the primary task. Note that our algorithm only uses the labels corresponding to the primary task ($y$) and does not require the labels for the secondary task ($z$) to provide privacy.

\section{Experiments}
ANI offers improved privacy during ML inference by injecting noise to the input on the client-side before transmitting it to the service provider. 
We perform experimental evaluations using various image classification tasks and report the trade-off offered by ANI between utility and privacy (by varying $\lambda$). Additionally, we compare our results with a prior work \emph{Cloak} and show that owing to the adaptive nature of our noise-injection scheme, ANI offers an improved trade-off between utility and privacy.

\begin{table}[htb]
\centering
\begin{tabular}{|l|l|l|}
\hline
\textbf{Dataset} & \textbf{Primary Task} & \textbf{Secondary Task} \\
\hline\hline
MNIST            & \textless 5 ?         & 10 class                \\
FashionMNIST     & \textless class 5 ?   & 10 class                \\
CIFAR-100        & 20 superclass         & 100 subclass               \\
CelebA-1           & Smiling ?             & Male / Female           \\
CelebA-2           & Smiling ?             & Young / Old          \\
\hline
\end{tabular}
\vspace{0.05in}
\caption{Datasets and the corresponding primary/secondary tasks used in our experiments to evaluate utility vs. privacy trade-off.}
\label{table:datasets}
\end{table}

\subsection{Datasets, Model and Training Parameters}

\textbf{Datasets:} We report the utility vs. privacy trade-off offered by ANI by performing evaluations on the image classification datasets mentioned in Table~\ref{table:datasets}. For each dataset, we consider two classification tasks-- a primary and a secondary task. The client is interested in performing remote inference for the primary task, while ensuring that the service provider is not able to infer the secondary (sensitive) attribute from the input. \texthl{For the CelebA dataset specifically, we consider two secondary tasks with the same primary task (CelebA-1, CelebA-2) to understand how the choice of secondary task impacts the utility-privacy trade-off.}

\textbf{$f_\theta$ architecture:} $f_\theta$ denotes the DNN used by the service-provider to perform inference on the primary task. We use a 5-layer Convolutional Neural Network (Conv-5) for MNIST and FashionMNIST~\cite{fashion} and a WideResNet16-4~\cite{wideresnet} model for CIFAR-100~\cite{cifar} and CelebA~\cite{celeba} datasets. Training is done using the SGD optimizer with an initial learning rate of 0.1 decayed with cosine annealing. We train the MNIST and FashionMNIST models for 50 epochs and the rest of the models for 100 epochs.

\begin{table}[htb]
\centering
\begin{tabular}{|m{0.2\columnwidth}|m{0.65\columnwidth}|}
\hline
\textbf{Network} & \textbf{Architecture}\\                                                                                          \hline\hline                                                                                         
WConv-2 &Conv 3x3, 8 - Maxpool / 2 - Linear - Upscale x 2 - Conv 3x3, 1 - Exp - Tanh \\
WConv-4 &Conv 3x3, 8 - Maxpool / 2 - Linear - Upscale x 2 - Conv 3x3, 1 - Exp - Tanh \\
\hline
\end{tabular}
\vspace{0.05in}
\caption{Model architectures used for $f_\psi$}
\label{table:f_w_networks}
\end{table}

\begin{figure}[htb]
	\centering
    \centerline{\epsfig{file=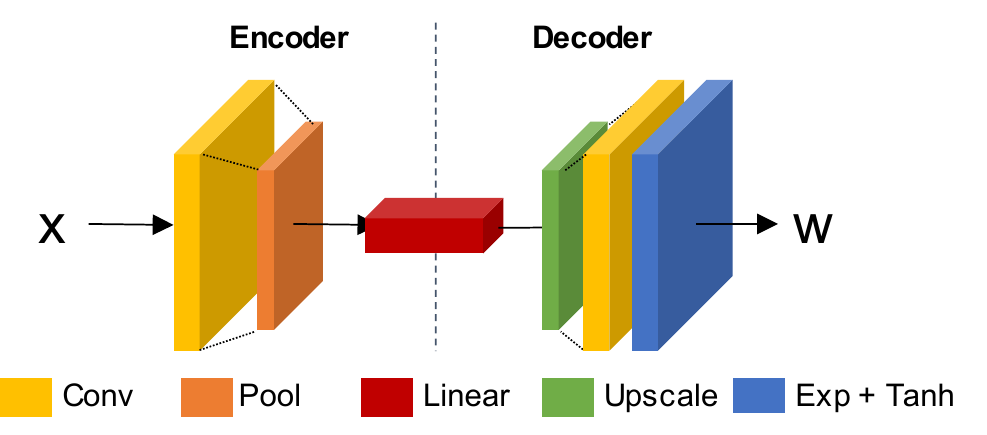, width=0.9\columnwidth}}
	\caption{Network architecture for WConv-2 used as $f_\psi$. The encoder reduces the spatial dimensionality to produce an embedding and the decoder increases the spatial dimensionality to produce a weight vector $w$ that matches the size of the input $x$.}
    \label{fig:wconv}
\end{figure}

\begin{figure*}[tb]
	\centering
    \epsfig{file=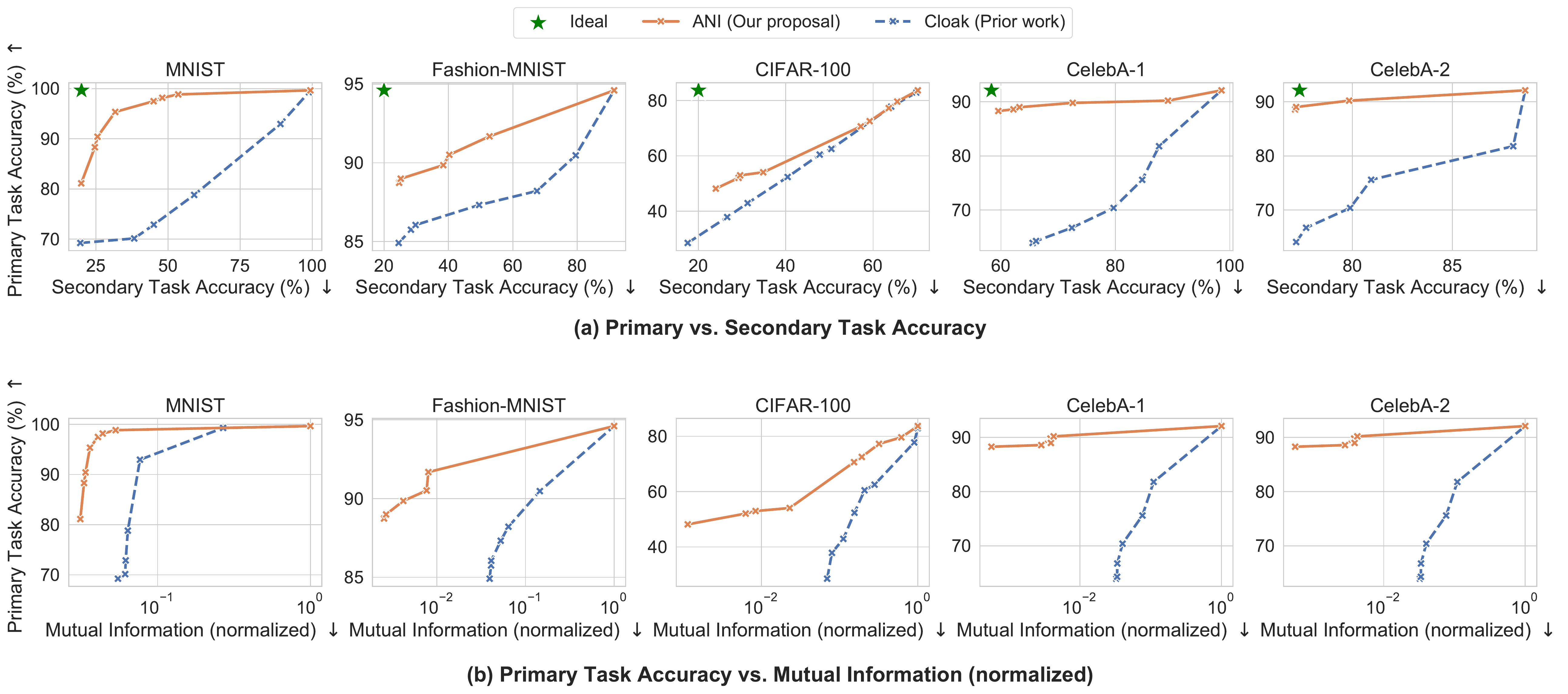, width=1.0\textwidth}
	\caption{Utility vs. Privacy trade-offs comparing ANI (our proposal) and Cloak (prior work). We consider two measures of privacy and report the corresponding trade-off curves: (a) Primary vs. Secondary Task Accuracy (b) Primary Task Accuracy vs. Mutual Information (normalized). Ideally we want to maximize utility (maximize primary task accuracy) and maximize privacy (minimize secondary task accuracy/mutual information). For both measures of privacy, ANI offers a significantly better trade-off between utility and privacy compared to Cloak across various image classification tasks.}
    \label{fig:results}
\end{figure*}

\textbf{$f_\psi$ architecture:} $f_\psi$ represents the DNN that outputs the weight vector $w$, which is used by the client to inject noise to the input, as described by Eqn.~\ref{eq:x'}. We use an encoder-decoder architecture for $f_\psi$ (similar to autoencoder networks~\cite{autoencoder}). The encoder reduces the spatial dimensionality of the input using convolution and pooling layers to produce an embedding. The decoder scales up this embedding using convolution and upsampling layers to produce a weight vector that matches the dimensions of the input. Fig.~\ref{fig:wconv} shows an example of such an encoder-decoder network. Table~\ref{table:f_w_networks} lists the architecture of the different encoder-decoder networks used in our evaluations. We use an $Exponent()$ followed by a $Tanh()$ layer at the end of each network to ensure that the weight values produced are in the range $[0,1]$ as required by our design. We use the WConv-2 network for MNIST and FashionMNIST and WConv-4 for CIFAR-100 and CelebA. We specifically chose $f_\psi$ networks that are small, to ensure low computational overheads on the client. However larger networks can also be used to provide a higher degree of customization for noise injection to obtain better utility vs. privacy trade-offs. We refer the reader to Appendix~\ref{app:compute} for a more detailed discussion on the computational overheads of $f_\psi$.

\subsection{Measuring Utility and Privacy}\label{sec:measuring_privacy}
We train the $f_\theta$ and $f_\psi$ models by choosing different values of $\lambda$ and report the utility and privacy offered by our scheme at each value of $\lambda$.

\textbf{Utility:} We measure utility by reporting the accuracy on the primary classification task obtained by evaluating the $f_\theta$ network on an unseen test set. 

\textbf{Privacy:} We measure privacy using two metrics:
\begin{enumerate}
    \item \textit{Secondary Task Accuracy:} We train a model $f_\phi$ on the secondary task and report the accuracy obtained by this model on a test set. This is a direct measure of the service provider's ability to infer the sensitive attributes of the client's input data and hence is inversely related to the privacy offered by the noise injection scheme.
    For a data point $(x_i,z_i)$, we train $f_\phi$ by first obtaining the noisy version of the inputs: $x'_i=f_\psi(x_i)$ and minimizing the cross-entropy between the predictions and the true labels: $\min_{\theta_s}\mathcal{H}(z_i,f_\phi(x'_i))$. The training parameters (learning rate, optimizer, epochs) used to train $f_\phi$ are the same as the ones used to train $f_\theta$. 
    \item \textit{Mutual Information:} ANI injects noise to the input $X$ to yield a noisy representation of the input $X'$ by removing non-conducive features from the input. The efficacy with which the information content is removed can be measured by computing the Mutual Information (MI) between $X$ and $X'$ denoted as: $\mathcal{I}(X, X')$. A smaller value of $\mathcal{I}(X, X')$ means that $X'$ contains a lower amount of information corresponding to the original input $X$ and hence provides a higher degree of privacy. We measure MI using the ITE Toolbox~\cite{ite}, which lets us estimate $\mathcal{I}(X, X')$ by collecting pairs of samples $(x_i,x'_i)$ using inputs from the test set.
    
\end{enumerate}

\subsection{Results}
Fig.~\ref{fig:results} shows the utility-privacy trade-offs offered by ANI, considering two measures of privacy (secondary task accuracy and mutual information) for the classification tasks listed in Table~\ref{table:datasets}. We compare the trade-off offered by ANI with that of a recent input-preprocessing-based proposal \emph{Cloak} to demonstrate the benefits of using adaptive noise injection over Cloak's fixed noise injection scheme.


\textbf{Primary vs. Secondary Accuracy:} Fig.~\ref{fig:results}a shows the trade-off between primary and secondary task accuracies. Ideally, we want our noise injection scheme to provide a high primary accuracy to maximize utility and a low secondary accuracy to maximize privacy. Our results show that, for a given primary accuracy (utility), ANI offers a significantly lower secondary accuracy (better privacy) compared to Cloak, resulting in a better utility-privacy trade-off that is closer to the ideal trade-off point. E.g., for the MNIST dataset, with a primary accuracy of $98\%$, the service provider can infer the sensitive attributes with an accuracy of just $48\%$ with ANI. In contrast, for the same primary accuracy, the service provider can infer sensitive attributes with a much higher accuracy of $97.8\%$ with Cloak.  \texthl{Additionally, our results for CelebA-1 and CelebA-2 show that ANI can provide good privacy regardless of the choice of the secondary (sensitive) attributes.}


\textbf{Primary Accuracy vs. Mutual Information:} Fig.~\ref{fig:results}b shows the trade-off between primary task accuracy and mutual information between $X$ and $X'$. We normalize the mutual information  $\mathcal{I}(X,X')$ with the entropy of the input $H(X)$. 
The goal of our noise injection scheme is to reduce $\mathcal{I}(X,X')$ to improve privacy while providing a high primary classification accuracy to maximize utility. Our results show that, for a given primary classification accuracy (utility), the mutual information $\mathcal{I}(X,X')$ produced by ANI is significantly lower than Cloak across all the datasets considered in our experiments. This demonstrates that ANI provides better privacy by removing a larger amount of non-conducive features from the input compared to Cloak.

\subsection{Visualizing Preprocessed Inputs}
\begin{figure}[tb]
	\centering
    \epsfig{file=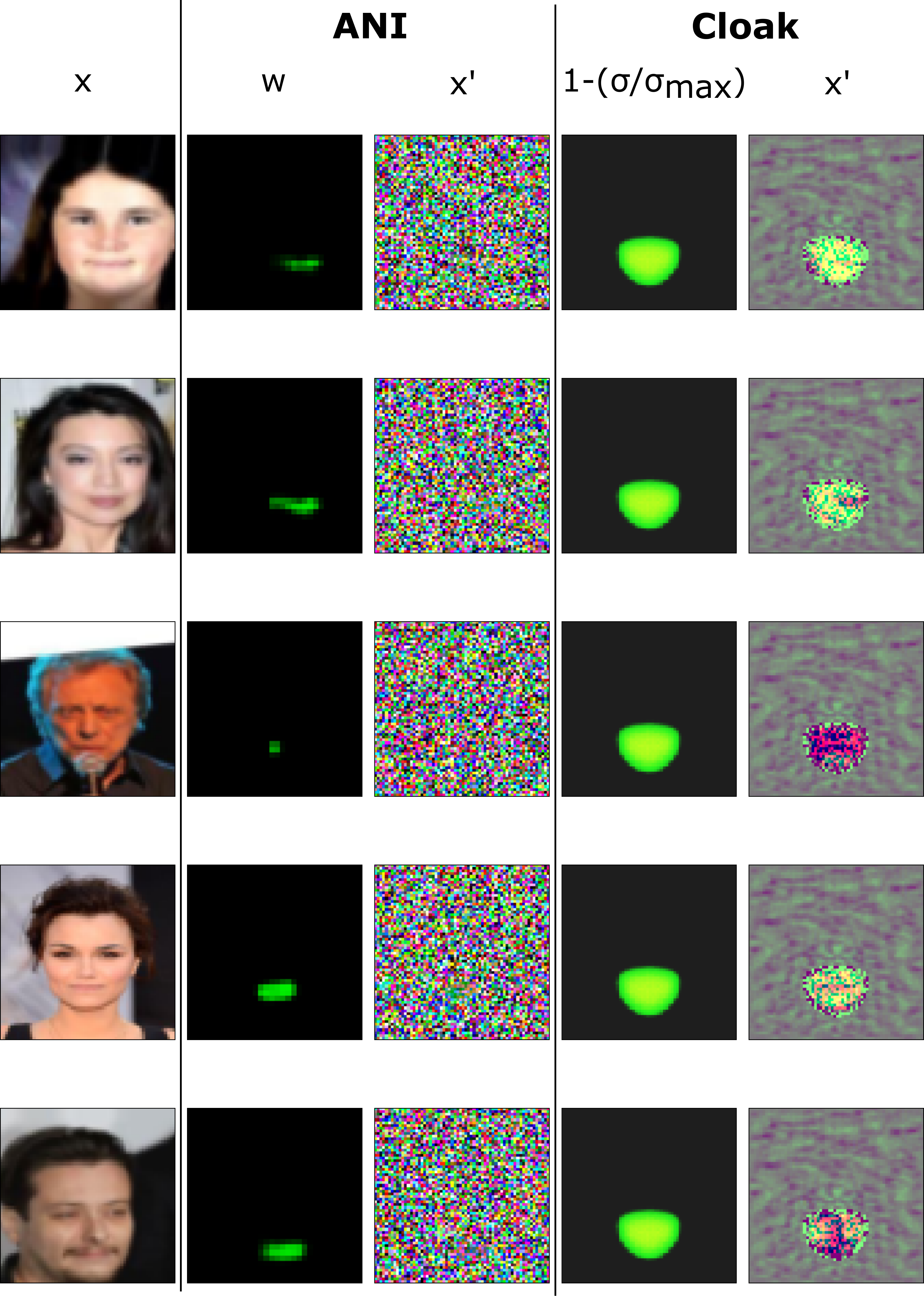, width=0.72\columnwidth}
	\caption{Comparing the noise masks produced by ANI with a recent prior work (\emph{Cloak}) for the smile detection task with CelebA dataset. ANI produces noise masks (weights $w$) that are tailored to the input, enabling ANI to obfuscate a greater fraction of the input. In contrast, Cloak produces a fixed noise mask, independent of the input. This adaptivity to the input enables ANI to achieve a better utility-privacy trade-off compared to Cloak.}
    \label{fig:transformed_inputs}
\end{figure}

Fig.~\ref{fig:transformed_inputs} visualizes the weights $w$ and the transformed inputs $x'$ produced by ANI for various inputs $x$ from the CelebA dataset for the smile detection task. Additionally, we visualize the transformed inputs $x'$ produced by Cloak as a point of comparison. Note that the spatial features of the input $x$ corresponding to high values of $w$ and $1-(\sigma/\sigma_{max})$ are preserved in the transformed representation $x'$. We make the following key observations:

\textbf{Hiding non-conducive features:} For the smile detection task, ANI produces a high value of the weights only near the mouth and low values everywhere else along the spatial dimensions of the image. Additionally, ANI chooses to retain information only along a single color channel (green) and hides the information in all the other color channels. By producing weight values that hide irrelevant parts of the input, ANI is able to produce a transformed representation $x'$, which is stripped of the information that is not pertinent to the primary classification task.

\textbf{Adaptivity to the input:} For the smile detection task, we ideally want to retain information closer to the mouth region. However, there can be significant variations in the spatial coordinates where the mouth of the person is found. The adaptive nature of ANI allows our method to customize the weights to capture this spatial variation and retain information corresponding to the mouth for each input image. In contrast, the prior work \emph{Cloak}, which uses a fixed noise distribution, has to resort to retaining a larger amount of spatial information to capture this variation, resulting in a worse utility-privacy trade-off compared to ANI.  

\section{Conclusion}

Users of remote machine learning inference services lack control over the computations that are performed on their data once it is sent to the service provider. This opens up the possibility for a malicious service provider to perform additional inference tasks that reveal sensitive attributes of the data without the user's consent, violating the user's privacy. This paper proposes \emph{Adaptive Noise Injection} (ANI), a noise injection scheme that improves inference privacy by obfuscating irrelevant features of the input. 
By varying the amount of noise, our scheme allows the user to trade-off utility for privacy offering a reduction of up to $\uptox\%$ in the sensitive task accuracy for $<1\%$ reduction in primary task accuracy. Owing to the input-specific nature of our noise injection scheme, ANI significantly outperforms prior work by offering a better utility-privacy trade-off.

{\small
\bibliographystyle{ieee_fullname}
\bibliography{egbib}
}
\newpage
\appendix
\section{Proofs}
\subsection{Upper bound on $\boldsymbol{\mathcal{I}(X'; Z)}$} ~\label{app:info_xy}

\textit{Theorem.} Given random variables  $X, X'$ and $Z$, where $X' = f(X)$, $\mathcal{I}(X'; Z)$ has the following upper-bound:

\begin{align} \nonumber
\mathcal{I}(X'; X) \geq \mathcal{I}(X'; Z)
\end{align}

\textit{Proof.} Consider $\mathcal{I}(X'; X,Z)$, which can be expressed in the following ways by the application of chain rule:

\begin{align} \label{eq:chain_1}
\mathcal{I}(X'; X,Z) = \mathcal{I}(X'; X) + \mathcal{I}(X'; X|Z) \\
\mathcal{I}(X'; X,Z) = \mathcal{I}(X'; Z) + \mathcal{I}(X'; Z|X) \label{eq:chain_2}
\end{align}

Since $X'$ is independent of $Z$ given $X$, we have $\mathcal{I}(X'; Z|X)=0$. Consequently, by equating Eqn.~\ref{eq:chain_1} and Eqn.~\ref{eq:chain_2}, we have:

\begin{align}\nonumber
\mathcal{I}(X'; X) + \mathcal{I}(X'; X|Z)= \mathcal{I}(X'; Z) \\
\implies \mathcal{I}(X'; X) \geq \mathcal{I}(X'; Z)\nonumber
\end{align}

\section{Computational Overheads of ANI}~\label{app:compute}
It is desirable to have low computational overheads for the network $f_\psi$ that is deployed on the client-side in ANI. This is because client devices like smartphones or smart home devices are likely to have low compute capabilities. To meet this requirement we designed the $f_\psi$ network to have low computational overheads. For reference, $f_\psi$ only requires $5\%$ of the FLOPs required to evaluate $f_\theta$ for MNIST and FashionMNIST. This number is even lower at $<0.5\%$ for CIFAR-100 and CelebA datasets.

\section{Validating the Privacy of ANI}~\label{app:training_privacy}
ANI proposes to train a network $f_\psi$, which is deployed on the client-side to inject noise to the input . $f_\psi$ is trained jointly with the network $f_\theta$ that is deployed on the server-side to perform the primary classification task. Since the training data is necessary to train both models, we expect the service provider to train $f_\psi$ and $f_\theta$. Once the service provider supplies the client with the noise injection network $f_\psi$, a natural question to ask  is: how does the client validate that $f_\psi$ indeed provides privacy as claimed by the service provider? We expect this validation to be done either by the client or by the open source community by evaluating the privacy metrics: 1. \emph{Secondary Task Accuracy} 2. \emph{Mutual Information}, using the $f_\psi$ network as described in Section~\ref{sec:measuring_privacy}. We expect this evaluation to be performed using open source datasets with a distribution that is similar to the client's data.

\section{Other Related Works}~\label{app:other_related_works}
In this section we discuss other related works that also use input-preprocessing to provide inference privacy and discuss their limitations.
Shredder~\cite{shredder} is a recent work which aims to provide inference privacy by evaluating the initial layers of a pre-trained neural network on the client-side and injecting noise to the intermediate representation before transmitting it to the service provider. Shredder uses a heuristically determined fixed noise distribution to inject noise for each input. The authors of Cloak (our baseline) show that their work provides a better utility-privacy trade off compared to Shredder. Thus, our proposal also outperforms Shredder due to the adaptive nature of our noise injection mechanism.

Leroux et al.~\cite{leroux2018privacy} aim to obfuscate the input image and make it unintelligible for a human observer using an autoencoder like architecture. They propose to do this by transmitting the embedding produced by the encoder network to the client instead of the input. While this method does make the transformed input unintelligible for a human observer, a malicious service provider would be able to train a model to infer sensitive attributes from such inputs, compromising the privacy of the client.

\end{document}